\documentclass[aip,jap,amsmath,amssymb,reprint,showpacs,floatfix,superscriptaddress]{revtex4-1}
\usepackage{graphicx}
\usepackage{dcolumn}
\usepackage{bm}
\usepackage{mathtools}
\usepackage{enumitem}
\usepackage{color}

\begin{document}


\title{Torque field and skyrmion movement by spin transfer torque in a quasi-2d interface in presence of strong spin-orbit interaction.} 



\author{Javier Osca}
\altaffiliation[New affiliation: ]{Department of Theoretical Physics, Maynooth University, Ireland.}
\affiliation{IMEC, Kapeldreef 75, B-3001 Leuven, Belgium}
\affiliation{KU Leuven, ESAT-MICAS, Kasteelpark Arenberg 10, 3001 Leuven, Belgium}
\author{Bart Sor\'ee}
\email{bart.soree@imec.be}
\affiliation{IMEC, Kapeldreef 75, B-3001 Leuven, Belgium}
\affiliation{KU Leuven, ESAT-MICAS, Kasteelpark Arenberg 10, 3001 Leuven, Belgium}
\affiliation{Universiteit Antwerpen, Departement Fysica, B-2000 Antwerpen, Belgium}

\date{Jul 7, 2021}

\begin{abstract}
We investigate the torque field and skyrmion movement at an interface between a ferromagnet hosting a skyrmion and a material with strong spin-orbit interaction. We analyze both semiconductor materials and topological insulators using a Hamiltonian model that includes a linear term. The spin torque inducing current is considered to flow in the single band limit therefore a quantum model of current is used.  Skyrmion movement due spin transfer torque proves to be more difficult in presence of spin orbit interaction in the case where only interface in-plane currents are present. However, edge effects in narrow nanowires can be used to drive the skyrmion movement and to exert a limited control on its movement direction. We also show the differences and similarities between torque fields due to electric current in the many and in the single band limits.
\end{abstract}


\maketitle 

\section{Introduction}
\label{intro}
\label{S1}
Moore's law for CMOS devices is reaching its limit and therefore spintronic technologies in general \cite{Scope1,Scope2} and skyrmions in particular\cite{Scope3} are considered as a way to overcome the constraints in power consumption and stability of CMOS devices. Furthermore, skyrmion movement using spin torque has attracted further attention because it has been proposed as one possible mechanism to build logic\cite{Processor1,Processor2} and memory devices\cite{Fert,Tomasello,Kang}.

In general, skyrmions are topological magnetic structures\cite{Topo} that arise when exchange and Dzyaloshinskii-Moriya interaction (DMI) are both present simultaneously. This may happen in the same material\cite{Bloch1,Bloch2,Bloch3} or in interfaces between two different materials\cite{Neel1,Neel2,Neel3}.  In the first case we obtain Bloch skyrmions while in the latter N\'eel skyrmions arise. In both cases the exchange interaction tries to minimize the energy of a ferromagnet aligning the spin of neighboring magnetic cells creating a uniform magnetization while DMI tries to minimize the energy, making neighboring spins to become perpendicular \cite{Dzyal,Moriya}. Skyrmions appear when the two mechanisms are balanced in their competition with each other. 

The main difference between bulk materials and interfaces is the plane in which DMI tries to make those spins perpendicular. This leads to different magnetic structures, where Bloch skyrmions show azimuthal magnetization in the skyrmion boundary while N\'eel skyrmions show a radial one\cite{Fert,Tomasello}.

There are two ways of achieving skyrmion movement, using spin transfer torque (STT) or using spin orbit torque (SOT)\cite{RevTorques}. Both torques are caused by the same mechanisms of exchange interaction between the spins of the bounded electrons responsible for the ferromagnet magnetization and the spin of conduction electrons flowing through the device\cite{STT1,STT2,STT3}. The main difference between the two is that interface STT implies a spin current moving in the interface plane while SOT implies spin current flowing perpendicularly to the interface plane and to the charge current parallel to it.

In this work we will focus ourselves on the study of N\'eel skyrmion movement in 2d interfaces because we want to study the interplay between magnetic topological structures and topological materials such as topological insulators and in general materials with strong linear terms in their Hamiltonians. Previous works have been focusing mainly on bulk materials capable of hosting the skyrmion by themselves or interfaces between a ferromagnet and heavy metals. These two materials (semiconductors and TI's ) imply a quantum mechanical model for the current because the more interesting properties of these materials arise in the regime of conduction by a single electronic mode. Very few works have considered quantum mechanical models for the current to calculate the spin torque \cite{QTopological,QuantumFourier,Osca3} in general or TI's in particular as in ref. \cite{Nagaosa} where an analytical approach was chosen where the effect of boundaries was not considered.  In this work we will also restrict ourselves to STT where the spin current flows in the interface plane. This way we can simplify the interface model to a 2d grid. More complicated 3d models for SOT are out of the scope of this work.

Both semiconductors and TI's have in common the presence of a linear term in their Hamiltonians resulting in spin accumulation on one of the nanowire edges. We will show how in both cases this accumulation has a dramatic impact on the nanowire electronic conductance and on the torque field affecting the skyrmion movement. Furthermore, there is an in plane spin-locking effect on the conduction electrons due the linear term that competes with the interaction between these electrons and the skyrmion structure. Consequently,  the consideration of many bands in topological insulators  will lead to different classical regimes of the torque in semiconductors with strong Rashba effect with respect to those of weak or zero spin-orbit interaction strength.

The paper is divided in five sections:
{\bf Section \ref{S1}: Introduction} introducing the topic and presenting some general concepts. {\bf Section \ref{S2}: Theoretical model} where the numerical method used in this work to calculate torque and skyrmion movement is described. {\bf Section \ref{S3}: Semiconductor results} where the torque field and skyrmion movement for semiconductor materials is studied. {\bf Section \ref{S4}: Topological insulator results} where the same is done for TI materials.  {\bf Section \ref{S5}: Multiband approach to the classical limit.} where we analyze for both semiconductors and TI cases where current is transmitted through the device by many bands. {\bf Section \ref{S6}: Conclusions} where key results are summarized.

\section{Model}
\label{S2}
\begin{figure}
  \resizebox{\columnwidth}{!}{
  \includegraphics{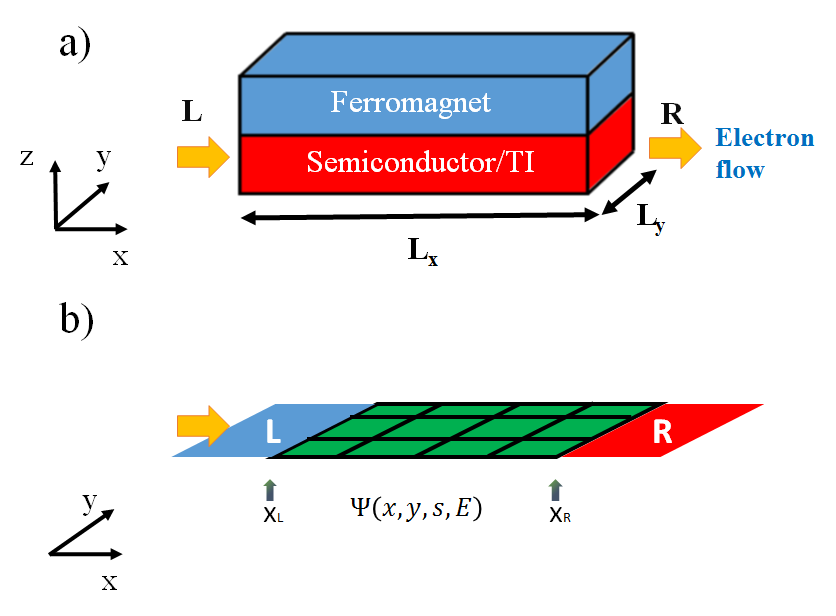}
  }
  \caption{ a) Schematic of a device holding a ferromagnet-semiconductor or alternatively a ferromagnet-TI interface. The skyrmion is held by the ferromagnet magnetization but its movement is driven by the torque created by the conduction electrons in the semiconductor or TI material. b) Quantum states of the conduction electrons are calculated in a 2d grid that represent the interface between the two materials. We consider the left and right boundaries to be open where in-going and outgoing states are modeled as plane waves. On the other hand, the upper and lower boundaries will be considered to be closed by a wall of infinite potential or to be periodic seaming the upper and lower edges depending on the case to be analyzed.}
  \label{F1}
\end{figure}

In this work we model the interface between a ferromagnet and a semiconductor or TI nanowire as a 2-dimensional planar interface as shown in fig. \ref{F1}. We consider that ferromagnet magnetization field support the skyrmion while an electron current flows through the nanowire.  Both effects coexist simultaneously in the interface where the spin of the conduction electrons interact with the skyrmion magnetization through STT.  To model the dynamics of the skyrmion in the ferromagnet we use the Landau-Lifshitz (LL) equation with an extra term to the effective field to account for the additional STT effect. This is,
\begin{equation}
\begin{split}
\frac{d{\bf M(r)}}{dt}=&-\frac{\gamma\mu_0}{1+\alpha^2} \left( {\bf M(r)} \times {\bf H}({\bf r}) \right)\\ &
-\frac{\alpha\gamma\mu_0}{1+\alpha^2} \left({\bf M(r)} \times  \left( {\bf M(r)} \times {\bf H}({\bf r}) \right) \right) 
\end{split}
\label{E16}
\end{equation}
where ${\bf M}({\bf r})=(M_x({\bf r}),M_y({\bf r}),M_z({\bf r}))$ is the ferromagnet magnetization, $\gamma$ is the gyromagnetic factor, $\mu_0$ is the vacuum permeability, $\alpha$ the dissipation coefficient and
\begin{equation}
{\bf H}({\bf r})={\bf H}_{\rm Z}({\bf r})+{\bf H}_{\rm ex}({\bf r})+{\bf H}_{\rm DMI}({\bf r})+{\bf H}_{\rm T}({\bf r})\,,
\label{E17}
\end{equation}
is the effective magnetic field felt by the magnetization. Note that as usual in the LL equation the magnetization magnitude is constant throughout the device $|{\bf M}({\bf r})|=M_{\rm s}$. There is one term in the effective magnetic field for each of the interactions present in the interface. Those terms are, the external Zeeman field ${\bf H}_{\rm Z}$ applied to the device, the exchange field responsible of the ferromagnetic properties of the material,
\begin{equation}
{\bf H}_{\rm ex}=\frac{2A}{\mu_0 M_{\rm s}^2} \nabla^2 {\bf M(r)}\,,
\label{E19}
\end{equation}
and the DMI effective field,
\begin{equation}
{\bf H}_{\rm DMI}=-\frac{2\mathcal{D}}{\mu_0 M_{\rm s}^2} \left(\left(\nabla \cdot {\bf M(r)} \right)\hat{z} - \nabla  M_z({\bf r}) \right)  \,,
\label{E21}
\end{equation}
where $A$ and $\mathcal{D}$ are the exchange and DMI constants respectively. We restrict ourselves to the use of perpendicular DMI because it is the one present in interfaces while parallel DMI is present in bulk materials. Finally, we also include the torque field, 
\begin{equation}
{\bf H}_{\rm T}=\frac{J_{\rm sd}}{\gamma_0 M_{\rm s}} {\bf m(r)}\,.
\label{E22}
\end{equation}
where $J_{\rm sd}$ is the exchange constant between the ferromagnet magnetization ${\bf M(r)}$ and the magnetization of the conduction electrons ${\bf m(r)}$.

Conduction electrons in the TI or semiconductor nanowire are modeled by the Hamiltonian,
\begin{equation}
\hat{\mathcal{H}}=\beta\, {\bf \hat{p}}^2+ \alpha \left(\hat{p}_x \hat{\sigma}_y - \hat{p}_y \hat{\sigma}_x \right) +{\bf \Delta}_B({\bf r}) \cdot {\bf \hat{\sigma}}\,,
\label{E23}
\end{equation}
where $\beta$ is the strength of the quadratic term, $\alpha$ the strength of the linear term and ${\bf \Delta}_{\rm B}({\bf r})=\frac{J_{\rm sd} S}{2} \frac{{\bf M(r)}}{M_{\rm s}}$. Note that the interaction between the ferromagnet and the conduction electrons is modeled as an external field where $J_{\rm sd}$ is the same exchange constant than in eq. \ref{E22} and $S$ is the dimensionless angular momentum of the pinned electrons in the ferromagnet, typically $S\approx2$. This is a general Hamiltonian that may be used to model both semiconductors and topological insulators with a different physical interpretation for each term. In a semiconductor $\beta=1/2m$ because the origin of the quadratic term is the kinetic energy of the conduction electrons while the linear term of strength $\alpha$ is the Rashba term that arises from the existence of an external electric field or from an effective field caused by growth anisotropies of the material. In this work we consider the second case therefore this field is pointing in the material growth direction ${\bf {\hat z}}$.  On the other hand, in a topological insulator the origin of the linear term in eq.\ref{E23} is the electron kinetic energy while the quadratic term arises as second order corrections to non-ideal materials\cite{Nagaosa}.

The device is modeled as a 2d grid shown in fig.\ref{F1}b where the contacts at the left and right sides of the device are modeled as open boundary conditions with known incident modes. Upper and lower boundaries may be a hard wall for a nanowire or  we can consider periodic boundary conditions for wider slabs. The magnetization of the conduction electrons  ${\bf m(r)}$  needed in eq. \ref{E22} is obtained from the integration of the magnetization of each occupied eigenstate of Hamiltonian in eq. \ref{E23},
\begin{equation}
\begin{split}
{\bf m}(x,y)=&-\gamma \frac{\hbar}{4\pi}\sum_{n_i}\int_{-\infty}^{\mu_L}{ \left(\,\Psi_{n_i}^*(E,x,y)\,{\bf \hat{\sigma}}\,\Psi_{n_i}(E,x,y)\,\right)}\,dE\\ 
             &-\gamma \frac{\hbar}{4\pi}\sum_{n_i}\int_{-\infty}^{\mu_R}{ \left(\,\Psi_{n_i}^*(E,x,y)\,{\bf \hat{\sigma}}\,\Psi_{n_i}(E,x,y)\,\right)}\,dE 
\end{split}
\label{E24}
\end{equation}
where  $\mu_L$ and $\mu_R$ are the chemical potentials of the left and right contacts respectively, $\Psi$ are eigenstates of eq. \ref{E23} and  ${\bf \hat{\sigma}}=\left({\bf \hat{\sigma}_{x}},{\bf \hat{\sigma}_{y}},{\bf \hat{\sigma}_{z}} \right)$ where ${\bf \hat{\sigma}_{x,y,z}}$ are the corresponding Pauli matrices.

Note that the occupation of the different eigenstates depend on the existence of an incident active mode in the left or right contact coupled with those states. Therefore their occupation depend on the chemical potential of the contacts. States active by the incidence of right going electrons $k>0$ depend on $\mu_L$ while the ones active by left going incident electrons $k<0$ depend on $\mu_R$. 

There is also the possibility of states attached to the interface with no input or output from the leads. If these states are pinned, therefore non-propagating, they will not be contributing to the torque because they must be aligned with the ferromagnet magnetization field.

\subsection{Resolution method.}
We calculate the time evolution of the ferromagnet magnetization ${\bf M}({\bf r},t)$ in the same way as in ref.\cite{Osca3}. The LL equation is numerically integrated to calculate the skyrmion movement as a function of time. A key point is that the dynamics of the ferromagnet magnetization is of a larger time scale than the conduction electrons dynamics. Therefore, at each time step of the LL integration the conduction electrons are assumed to be relaxed to their equilibrium states and we can use an adiabatic approximation to calculate their effect in the skyrmion dynamics. In this approximation, the eigenstates of the Hamiltonian equation \ref{E23} are used to obtain ${\bf m}({\bf r},t)$ with eq. \ref{E24}. The magnetization of the conduction electrons ${\bf m}({\bf r},t)$ is then further used to calculate the torque field of eq. \ref{E22} required for the LL integration.

In order to have an electron flux we consider contacts where the incident modes are modeled as plane waves, therefore the full solutions in these contacts are the linear superposition of the incident and reflected waves. This is for a given energy,
\begin{equation}
\label{E26}
\Psi^c(E,x,y,s) =
\sum_{\alpha,n_\alpha} 
{\frac{d^{(c,\alpha)}_{n_\alpha}}{\sqrt{\hbar v^{(c,\alpha)}_{n_\alpha}  }}
\exp{\left[ik^{(c,\alpha)}_{n_\alpha}x\right]}
\phi^{(c,\alpha)}_{n_\alpha}(E,y,s) 
}\; ,
\end{equation}
where $c=L,R$ labels the contacts, $\alpha=i,o$ the input and output modes in each contact,  $s=\uparrow,\downarrow$ the spin up and down, $k^{(c,\alpha)}_{n_\alpha}$ their wavenumber and
\begin{equation}
v^{(c,\alpha)}_{n_\alpha}=\frac{1}{\hbar}\frac{\partial E}{ \partial k^{(c,\alpha)}_{n_\alpha} }
\label{E26B}
\end{equation}
their group velocity . The coefficients $d^{(c,\alpha)}_{n_\alpha}$ will determine the amplitudes of the asymptotic solutions in the leads while  $k^{(c,\alpha)}_{n_\alpha}$ and $\phi^{(c,\alpha)}_{n_\alpha}$ are assumed known.
 
The eingenstates of the Hamiltonian eq. \ref{E23} are obtained in the same manner as in refs. \cite{Osca3,Osca1,Osca2}.  The Hamiltonian equation \ref{E27} is fulfilled in a central region where $E$ is used as a parameter while the matching eqs. \ref{E28} and \ref{E29} allow the consideration of open boundary conditions connecting this central region with the left and right contacts. The full system of equations is then solved numerically \cite{Harwell}, 
\begin{equation}
\label{E27}
\left(\hat{\mathcal{H}}-E\right) \Psi(E,x,y,s) = 0\; ,\qquad\;\,(xy)\in C \; ,	\;
\end{equation}
\begin{equation}
\label{E28}
\begin{split}
\Psi(E,x,y,s) 
-&
\sum_{n_o}
{\frac{d^{(c,o)}_{n_o}}{\sqrt{\hbar v^{(c,o)}_{n_o}  }}
\exp{\left[ik^{(c,o)}_{n_o}\right]}
\phi^c_{n_o}(E,y,s)
}
=\\ 
&\sum_{n_i}
{\frac{d^{(c,i)}_{n_i}}{\sqrt{\hbar v^{(c,i)}_{n_i}  }}
\exp{\left[ik^{(c,i)}_{n_i}\right]}
\phi^c_{n_i}(E,y,s)
}\; , \\
\,&\quad\quad\quad\quad\quad\quad\quad\quad\quad\, (x,y,c)\in L/R \; ,\quad\\
\end{split}
\end{equation}
\begin{equation}
\label{E29}
\begin{split}
&\sum_{s}\int{dy\,
\phi^{(c,o)}_{m_o}(E,y,s)^*
\, \Psi(E,x_c,y,s)}  
-\\
&\sum_{n_o}
{\frac{d^{(c,o)}_{n_o}}{\sqrt{\hbar v^{(c,o)}_{n_o}  }}
\exp{\left[ik^{(c,o)}_{n_o}x_c\right]}
{\cal M}_{m_o n_o}^{(o c,o c)}(E)
}
= \\ 
&\sum_{n_i}
{\frac{d^{(c,i)}_{n_i}}{\sqrt{\hbar v^{(c,i)}_{n_i}  }}
\exp{\left[ik^{(c,i)}_{n_i} x_c\right]}
{\cal M}_{m_o n_i}^{(o c,i c)}(E)
} \; ,
\; c\in L/R \; , \;
\end{split}
\end{equation}
where $x_c$ is the coordinate of the boundary $c=L,R$ with the central region and  
\begin{equation}
{\cal M}_{m_\alpha n_\beta}^{(\alpha c,\beta c)}(E)
=
\sum_{s}
\int{dy\,
\phi_{m_\alpha}^{(\alpha,c)}(E,y,s)^*
\phi_{m_\beta}^{(\beta,c)}(E,y,s)
}\; .
\end{equation}

In this work some of the numerical results will be given in characteristic units of the device where the energy and length units are $E_U=\hbar^2/mL_y^2$ and $L_U=L_y$ respectively.

\section{Ferromagnet-semiconductor interface}
\label{S3}
\subsection{Torque field and skyrmion movement}
\begin{figure}
  \resizebox{\columnwidth}{!}{
  \includegraphics{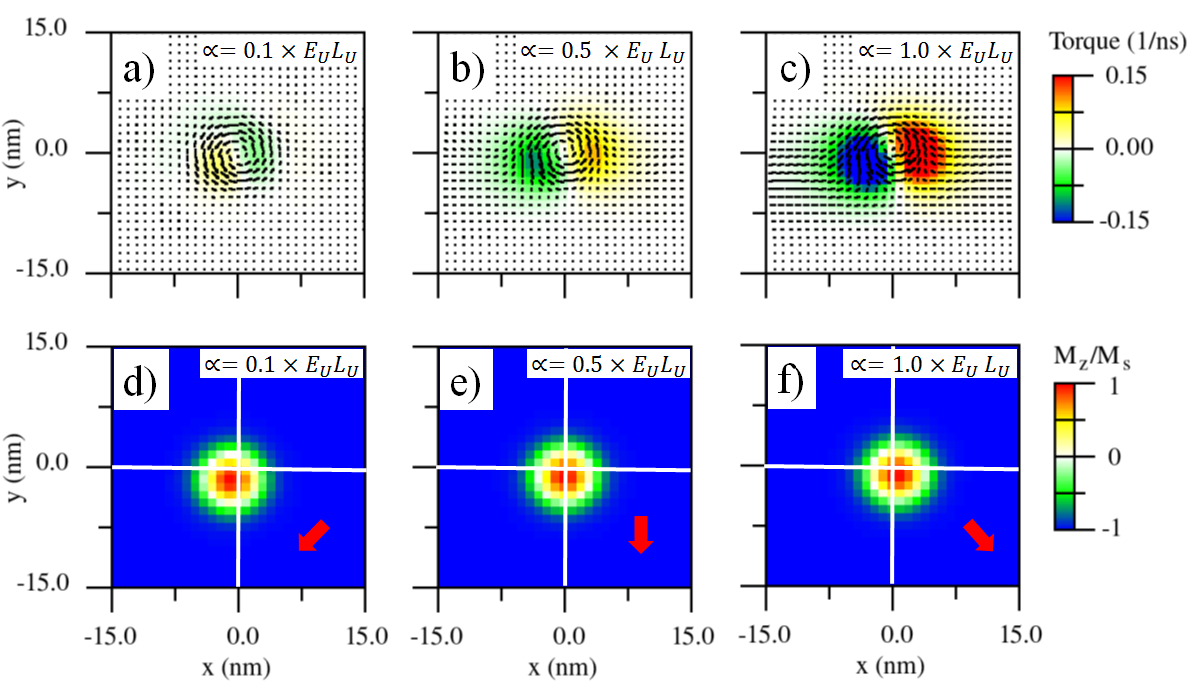}
  }
  \caption{ a) Torque field when a skyrmion is located at $x=y=0$ when the ferromagnet is in contact with a semiconductor of $\alpha=0.1\times E_UL_U\equiv 23.5\, {\rm meV\,nm}$. b) and c)  the same for devices with semiconductor spin-orbit strengths $\alpha=0.5\times E_UL_U\equiv 117.5\,{\rm meV\,nm} $ and $\alpha=1\times E_UL_U\equiv 235\,{\rm meV\,nm}$ respectively. d) $z$ component of the ferromagnet magnetization holding a skyrmion after $10ns$ simulation when the current causing the torque and therefore the skyrmion movement flows from left to right through a semiconductor of $\alpha=0.1\times E_UL_U$. The skyrmion original position $x=y=0$ is shown as the crossing point of the two white lines. e) and f) the same than in d) for devices where the semiconductor spin-orbit strength ore $\alpha=0.5\times E_UL_U$ and $\alpha=1\times E_UL_U$ respectively.  The rest of the parameters that characterize the interface are the same for all the figures in order to allow for a comparison. These are:  $J=400\;{\rm meV/nm}$, $D=500\;{\rm meV/nm^2}$, $\hbar\gamma_0 H_z=0.2\;{\rm meV}$,  $Jsd=45.0\;{\rm meV}$, $S=2$, $m/m_e=0.013$ and $\mu_L=75\;{\rm meV}$ where $m_e$ is the bare electron mass.}
  \label{F2}
\end{figure}

\begin{figure}
  \resizebox{\columnwidth}{!}{
  \includegraphics{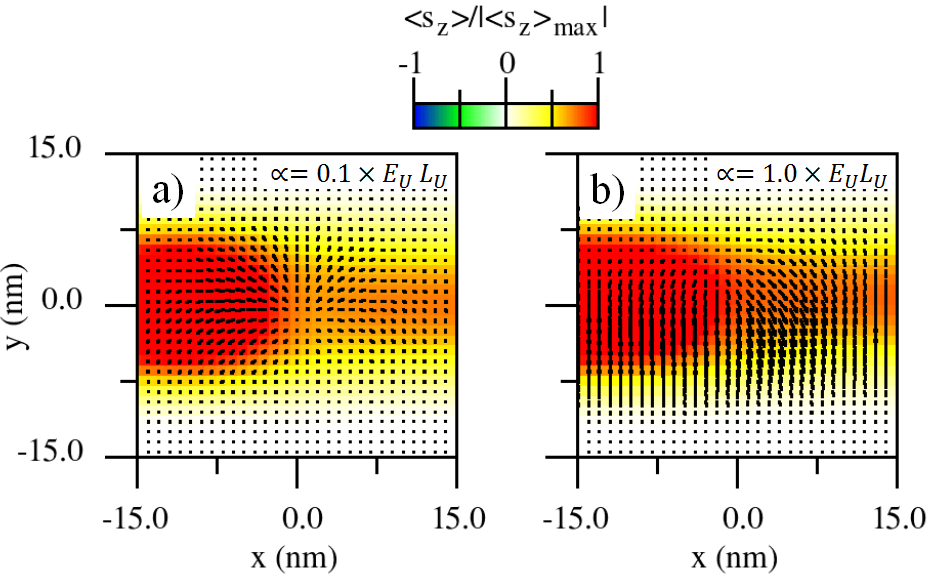}
  }
  \caption{ a) and b) Conduction electron magnetization for the same cases than in fig.\ref{F2}a and fig.\ref{F2}c respectively.  $z$ component of the magnetization is presented in color while $x$ and $y$ component are shown as vectors.}
  \label{F6}
\end{figure}

As explained earlier, in a semiconductor the physical origin of the quadratic term in eq. \ref{E23} is the kinetic energy of the electrons. Therefore $\beta=1/2\times E_U L_U^2$ while the linear term arises from Rashba spin-orbit interaction (SOI). In figures \ref{F2}a, \ref{F2}b and \ref{F2}c the torque field,
\begin{equation}
T=-\frac{1}{1+\alpha^2} \frac{J_{\rm sd}}{\gamma_0 M_{\rm s}}\left( {\bf M(r)} \times {\bf m(r)} \right)\,,
\label{E30}
\end{equation}
is visualized for three different values of the spin-orbit coupling strength $\alpha$. We can see that for a small $\alpha=\times 0.1\,E_UL_U$ the torque field is similar than the one reported for metals \cite{Osca3}. However, as we  increase $\alpha$, a spin-y accumulation builds up at the bottom side of the nanowire as shown in fig. \ref{F6}b for the particular case of $\alpha=1\times E_UL_U$.  Note that the build-up happens only at the bottom side because we are considering incident modes only at the left terminal. In equilibrium conditions a build up of spin-y of the opposite spin will also happen at the upper boundary. This accumulation affects the shape of the torque that acquires a non-zero value also at the bottom of the nanowire pointing in the $-\hat{x}$ direction due the spin-y accumulation. As expected, torque strength also increases with $\alpha$ as the spin-y accumulation increases. 

This change in the torque shape affects directly the torque movement. In figs. \ref{F2}d, \ref{F2}e and \ref{F2}f we can see the resulting skyrmion movement after $10\,{\rm ns}$ for different spin-orbit strengths. For $\alpha=0.1\times E_UL_U$ the spin-orbit effect is small enough to have little impact on the torque field and therefore the skyrmion movement is similar to the one obtained in metals with this same kind of model. In this case the torque asymmetry created by the electrons impinging from left to right make the skyrmion move in a perpendicular direction to the electron current ( a certain backwards component is also present whose origin is explained below). 

A more complicated process occurs when the Rashba term is dominant. For $\alpha=1\times E_UL_U$  the skyrmion moves in a left to right direction. In this regime the accumulation of spin at the bottom of the nanowire makes the torque more intense at the bottom than the top of the skyrmion thus creating an imbalance of forces that makes the skyrmion move in the longitudinal x-direction. This is similar to the phenomena reported in ref.\cite{Osca3} where skyrmions near nanowire edges move because the decay of the electron density and therefore also the spin density near those edges thus creating a similar kind of force imbalance. Both cases have in common that a change in the spin density of the conduction electrons lead to a change in the  torque field that leads to a non-zero balance of forces. However, the main difference is that here the spin accumulation is not correlated with the electron density distribution. In this case the maximum in the conduction electron magnetization ${\bf m(r)}$ does not coincide with the maximum in electron density.

Finally, in the intermediate regime $\alpha=0.5\times E_UL_U$ where the kinetic energy and the Rashba term have a similar strength both effects appear combined.
The resulting skyrmion movement therefore appears as an intermediate regime that combines features of the two previous cases. The perpendicular movement of the skyrmion with respect the electron flow when the strength of the spin-orbit coupling is small $\alpha=0.1\times E_UL_U$ displaces the skyrmion away from the nanowire center. When the skyrmion is no longer at the nanowire center the density at the top of the skyrmion becomes larger than at the bottom therefore the resulting torque field is also larger at the top creating some backward movement in this case. We can see that this effect is compensated in the intermediate regime $\alpha=0.5\times E_UL_U$  by the spin accumulation at the bottom of the wire resulting in a perfectly perpendicular to the wire skyrmion movement without any backward component.

\subsection{Current blockade and distribution}
\begin{figure}
  \resizebox{\columnwidth}{!}{
  \includegraphics{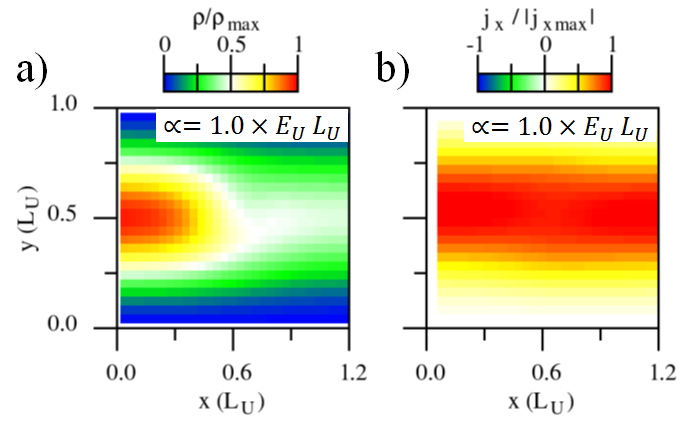}
  }
  \caption{ a) Conduction electron density for the same interface parameters than in fig. \ref{F2}c. b) $x$ component of the electron particle current for the same case than a). Note how the skyrmion presence increases electron density at its left and slightly shifts the current maximum. }
  \label{F4}
\end{figure}

\begin{figure}
  \resizebox{\columnwidth}{!}{
  \includegraphics{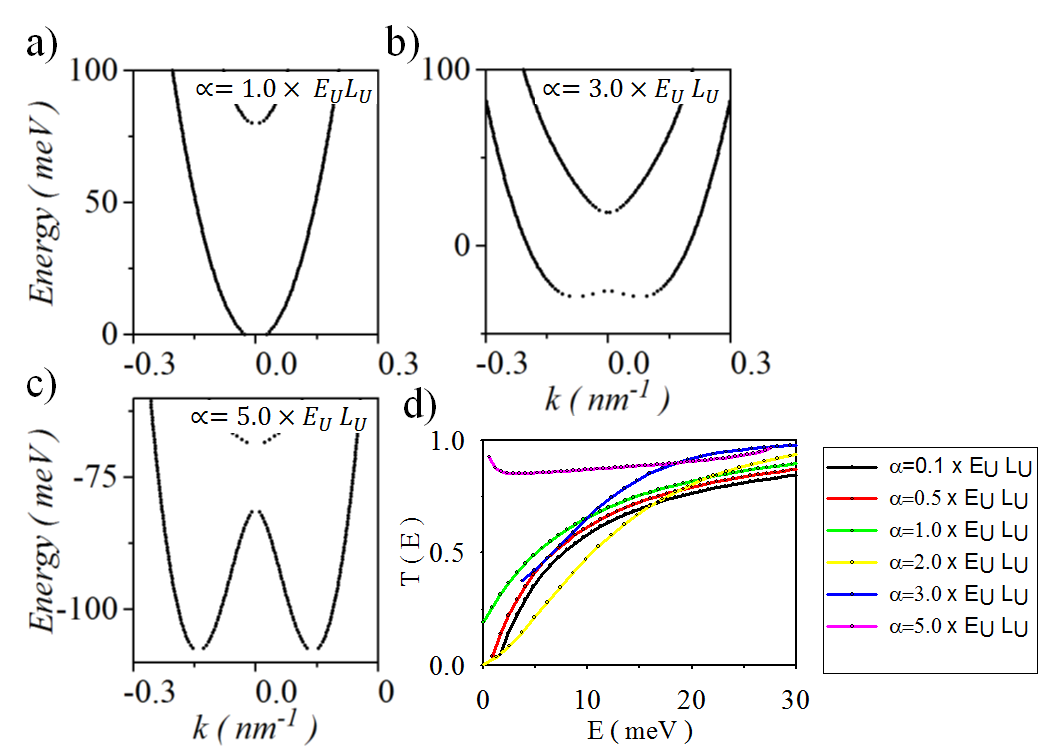}
  }
  \caption{ a) b) and c) spectrum of a semiconductor nanowire of spin-orbit constant strength $\alpha=1.0 \times E_UL_U$, $\alpha=3.0 \times E_UL_U$ and $\alpha=5.0 \times E_UL_U$ respectively. The rest of the parameters are the same than in fig. \ref{F2}. d) Transmission probability of a semiconductor nanowire for different values of the spin-orbit strength $\alpha$ as a function of the energy referred to the bottom of the band. The line for $\alpha=5.0\times E_U L_U$ (in purple) is the transmission probability of just one of two injected modes, this way allowing for a comparison.}
  \label{F4B}
\end{figure}

Conduction electrons perceive a skyrmion as a magnetic inhomogeneity that may block the flow of current.  We can see in fig. \ref{F4}a how the maximum of the electron density is at the left side of the skyrmion because injected electrons are impinging the skyrmion from the left contact. The effect of this blockade can also be seen in the conductance in fig. \ref{F4B}d where except in certain cases (more on this below) the current is blocked for energies near the bottom of the conduction band. This blockade already occurs in metals $\alpha=0\times E_U L_U$ \cite{Osca3} but new additional effects arise with increasing $\alpha$. 

\begin{figure}
  \resizebox{\columnwidth}{!}{
  \includegraphics{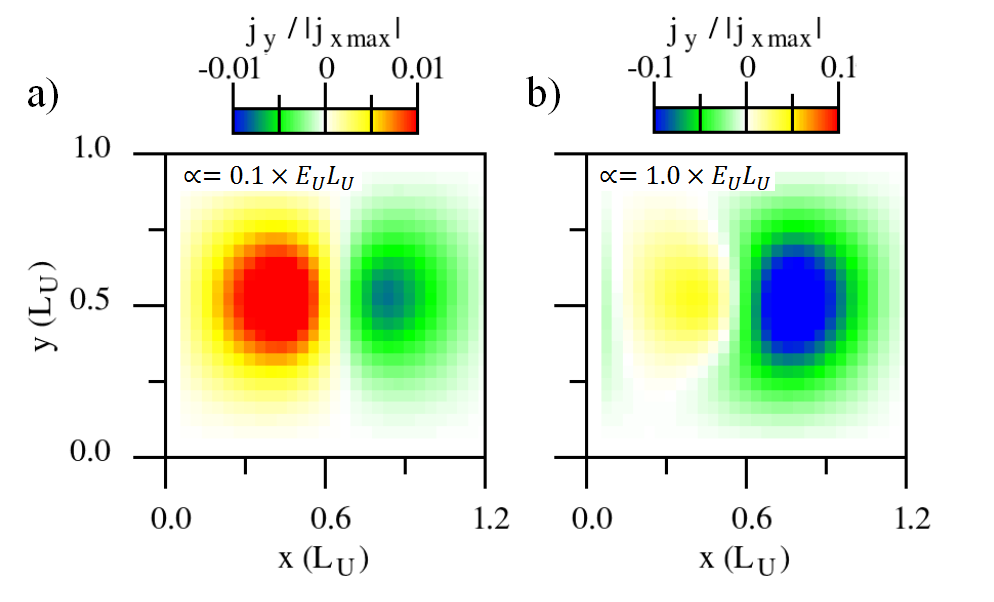}
  }
  \caption{ a) and b) y component of the electron particle current for the same case than fig. \ref{F2}a and \ref{F2}c respectively.}
  \label{F5}
\end{figure}

In fig. \ref{F4}b its shown how for a dominant linear term in the Hamiltonian $\alpha=1\times E_UL_U$ the longitudinal component of the current is shifted due to the presence of the skyrmion. The maximum of the $x$ component of the current is slightly shifted from $y=0$ to a small negative value at the right side of the skyrmion. This does not happen for $\alpha=0.1\times E_U L_U$ or for metals $\alpha=0\times E_U L_U$. However, the most notable change in the distribution of current can be seen in the $y$ component of the current as shown in fig. \ref{F5}. For a low value of the SOI constant $\alpha=0.1\times E_UL_U$ the current has a positive $y$ component at the left of the skyrmion and a negative one at the right similar in magnitude. Therefore we can assume some deflection of the electronic current when interacting with the skyrmion. However, for a large value of $\alpha=1\times E_UL_U$ the downward component at the right of the skyrmion is larger than the upward at the left therefore creating a recoil that explains the difference in the skyrmion movement direction with respect the case where $\alpha=0.1\times E_UL_U$.

The transmission probability dependence on different values of the spin-orbit interaction $\alpha$ in the nanowire material is not straightforward as can be seen in fig. \ref{F4B}d. Note that for the cases where there is only a single input mode conductance in the leads and transmission probability $T$ are related by the expression, $T(E)=g(E)/(e^2/h)$ where $E$ is the energy of the injected electrons. 

For values of the SOI constant between $\alpha=0.1\times E_UL_U$ and $\alpha=1.0\times E_UL_U$ the conduction bands in the contacts still follow a mainly parabolic dispersion relation with a small deformation due the Rashba SOI as plotted in fig. \ref{F4B}a.  In fig. \ref{F4B}d we can see how for this particular range of values of $\alpha$ the transmission probability profile with energy is approximately independent of $\alpha$ .  We can also see a transmission blockade due the skyrmion presence for energies near the bottom of the conduction band.

On the other hand, as can also be seen in fig \ref{F4B}b for values of the spin-orbit constant between $\alpha=2\times E_U L_U$ and $\alpha=3\times E_U L_U$ there are four propagating modes available for energy values near the bottom of the band. The overall transmission at the onset of the four mode regime $\alpha=2.0\times E_UL_U$ drops but increases quickly again for strengths of the SOI larger than $\alpha=3.0\times E_U L_U$. 

Finally, as shown in fig. \ref{F4B}c for larger values of SOI  $\alpha=5.0\times E_UL_U$ we have four propagating modes all along the single band regime. In this case we have chosen to show the transmission probability of the $k>0$ injected mode of larger wavenumber value. Note that the transmission probability of both $k>0$ input modes are similar in this particular scenario. Interestingly for large values of SOI in the nanowire the skyrmion becomes invisible to current transmission measurements.

In general, we believe that a more intense effective electric field locks more tightly the conduction electrons spins into the $y$ direction thus limiting the recoil that the conduction electrons will absorb and therefore diminishing the blockade. The external magnetic field that the skyrmion represents to the flowing electrons becomes less important than the effective field that the electrons perceive due to spin-orbit interaction. 

\section{Ferromagnet-TI interface}
\label{S4}
In an ideal TI the electron dispersion relation is linear. Therefore to model a TI the constant $\beta$ in eq. \ref{E23} must be zero. In this case the physical origin of the linear term must be reinterpreted as the kinetic energy of the electrons in the TI nanowire surface without need of any further interaction. However, it is not possible to model a TI nanowire  with upper and lower hard wall boundary conditions in this case because the only solution to this Hamiltonian with these boundary conditions is the zero solution. To avoid this problem there are two possible approaches. First, to consider periodic (or anti-periodic) boundary conditions in the upper and lower edges or to maintain the hard wall boundaries and consider the presence of second order quadratic correction to the Hamiltonian, eq. \ref{E23}, due to non-idealities in the material\cite{Nagaosa}. The first approach implies modeling wide slabs instead of a nanowire, neglecting edge effects while the second approach implies the use of a  non-zero coefficient $\beta<<1/2\times E_U L_U^2$ in eq. \ref{E23}. We have found no skyrmion movement using periodic boundary conditions (wide slab) for TI's if we consider only a linear term $\beta=0$. Skyrmion movement is obtained however if second order corrections are considered but this movement is similar to the one obtained with second order terms alone.

\begin{figure}
  \resizebox{\columnwidth}{!}{
  \includegraphics{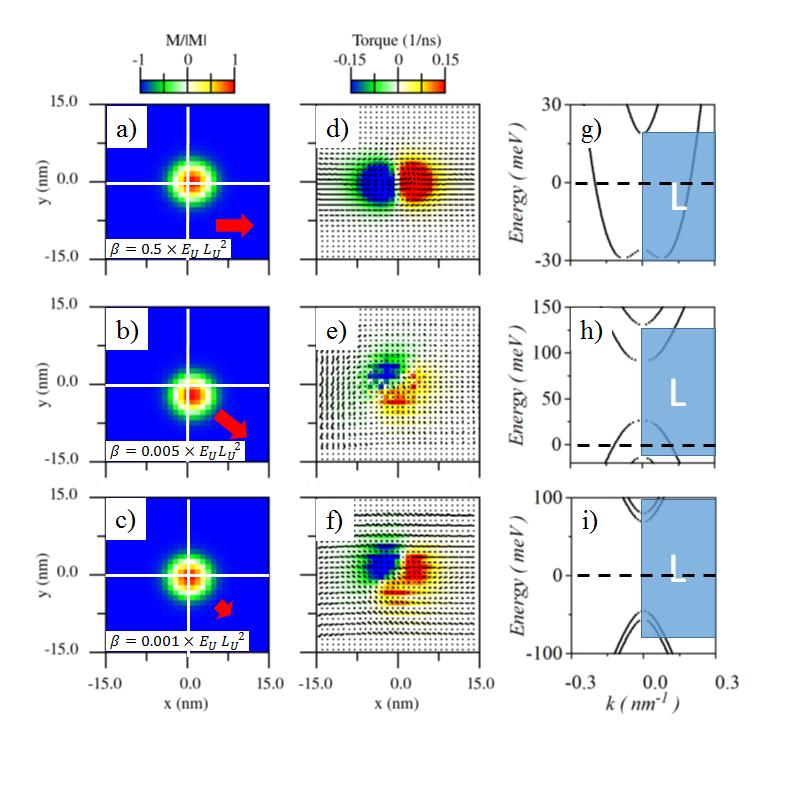}
  }
  \caption{ a) skyrmion position after a $10\,ns$ simulation when the electron current causing the skyrmion movement flows through a semiconductor nanowire $\beta=1/2\times E_U L_U^2$ with a large dominant spin-orbit strength $\alpha=3\times E_U L_U$ and  $m/m_e=0.013$ where $m_e$ is the bare electron mass, b) and c) the same than a ) for the case where the nanowire is made of a non-ideal topological insulator with second order quadratic correction of strengths $\beta=0.005\times E_U L_U^2$ and $\beta=0.001\times E_U L_U^2$ respectively. The value of the spin-orbit coupling strength is also $\alpha=3\times E_U L_U$ in both cases. d), e) and f)  are the corresponding torque fields to a),b) and c) while g),h) and i) are the spectra of the propagating bands in the contacts for each case. The rest of the parameters are also the same than in fig. \ref{F2}. These are:  $J=400\;{\rm meV/nm}$, $D=500\;{\rm meV/nm^2}$, $\hbar\gamma_0 H_z=0.2\;{\rm meV}$,  $Jsd=45.0\;{\rm meV}$, $S=2$ and $\mu_L=75\;{\rm meV}$. Note that $E_U=10\;{\rm meV}$ for both the semiconductor and the TI for comparison purposes.}
  \label{F7}
\end{figure}

On the other hand, hard wall boundary conditions for the upper and lower edges (nanowire) are more interesting. Similar to the previous section, the linear term implies a spin-y accumulation in the lower nanowire edge but now $\beta<<1/2\times E_U L_U^2$ therefore electron propagation in the nanowire is different than in semiconductors and consequently the torque field and skyrmion movement direction are also different. In figs. \ref{F7}a, \ref{F7}b and \ref{F7}c the resulting skyrmion movement by STT caused by current flowing through three different nanowires with three different values of $\beta$ is shown. In figs. \ref{F7}d, \ref{F7}e and \ref{F7}f it is also shown their corresponding torque fields and in figs. \ref{F7}g, \ref{F7}h and \ref{F7}i we have plotted the bands in the contacts of the incident conduction electrons also for the same three cases.

A smaller $\beta$ implies that the quadratic term is less relevant compared to the linear term and therefore the spectrum evolves from a typical Rashba-like spectrum (see fig. \ref{F7}g) to a typical TI spectrum as plotted in fig. \ref{F7}i.  In the calculation of the conduction electrons magnetization ${\bf m(r)}$ for TI we restrict the lower bound of the integration in eq. \ref{E24}  to a finite value under the assumption that the magnetization interaction between the ferromagnet and the conduction electrons diminish for electrons deep in the lower energy band. This minimum value is the smallest energy value plotted in figs. \ref{F7}g and \ref{F7}i .Furthermore, in these two figures it is also only plotted the portion of the bands close to $k=0$. Those same bands close the gap for large $|k|>0$ wavenumbers but these external branches are dismissed in the calculation of ${\bf m(r)}$ for TI (and not shown in figs. \ref{F7}g and \ref{F7}i) limiting the integration up to a maximum wavenumber. We are assuming an input electron flow
only in the inner branches and we consider scattering to the external branches of very large wavenumber unlikely.

The necessary magnetic interaction between the skyrmion and the conduction electrons may open a significant gap at $k=0$. In the case where $\beta$ is close to zero ($\beta=0.001 \times E_U L_U^2$) this interaction implies losing the linear dispersion close to $k=0$. For physically reasonable parameters we can see in figures \ref{F7}c, \ref{F7}f and \ref{F7}i how the spin up and down bands are close enough in a way that the overall spin and spin torque is almost canceled therefore canceling skyrmion movement. A smaller interaction between the skyrmion and the conduction electrons would imply to recover a linear or quasi-linear dispersion relation but it may not generate enough torque to create skyrmion movement neither.

We can also see in figs. \ref{F7}a, \ref{F7}d and \ref{F7}g the results where we set the quadratic term to $\beta=1/2\times E_U L_U^2$ like semiconductor materials (but for larger SOI strength than in the previous section) for comparison purposes. In this case the spin-y accumulation in the lower edge combines with the torque asymmetry to move the skyrmion from left to right in the way already discussed in the previous section. 

\begin{figure}
  \resizebox{0.75\columnwidth}{!}{
  \includegraphics{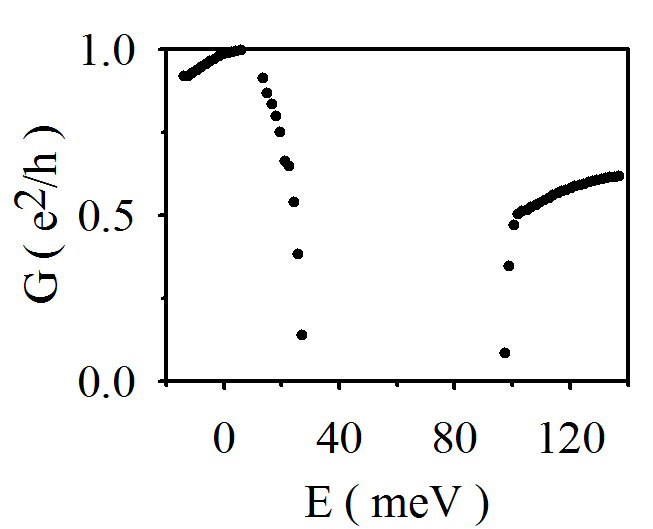}
  }
  \caption{d) Conductance of a topological insulator nanowire as a function of the energy for the same parameters than in fig. \ref{F7}b.}
  \label{F8}
\end{figure}

Finally, as shown in fig. \ref{F7}h, it is possible to attain spin polarized bands for an intermediate regime $\beta=0.005\times E_U L_U^2$ where 
their energy separation with respect the bottom of the bands is larger than in previous cases but at the expense of larger band deformations with respect a linear dispersion relation. In this case the interplay between  conduction electrons and the skyrmion becomes more complicated and we obtain a diagonally moving skyrmion movement (see fig \ref{F7}b). This regime is interesting from the point of view of skyrmion detection. In fig. \ref{F8} the transmission probability drop is shown near the bottom of the upper band (like in metals or semiconductors) and a similar drop at the top of the lower band. Interestingly these conductance drops are asymmetric with respect to the band gap at $k=0$ and therefore they are different for the top and bottom bands.

\section{Multiband approach to the classical limit.}
\label{S5}
\begin{figure}
  \resizebox{\columnwidth}{!}{
  \includegraphics{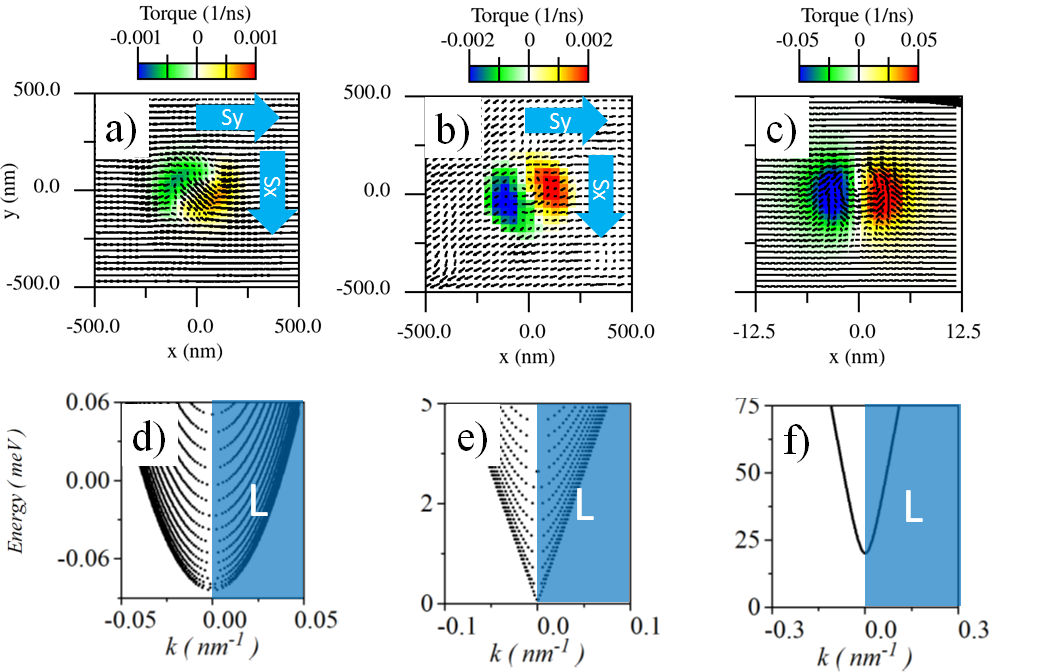}
  }
  \caption{ a) Torque field resulting from the interaction with a current carried out by multiple bands of a semiconductor $\beta=1/2\times E_U L_U^2$ where $\alpha=10\times E_U L_U$ using hard wall boundary conditions for the upper and lower boundaries. The rest of the parameters for the semiconductor material are are:  $J=125\;{\rm meV/nm}$, $D=25\;{\rm meV/nm^2}$, $\hbar\gamma_0 H_z=0.0004\;{\rm meV}$,  $Jsd=0.1\;{\rm meV}$, $S=1.7$, $m/m_e=0.53$ and $\mu_L=75\;{\rm meV}$ where $m_e$ is the bare electron mass. The resolution of the numerical discretization is: $\Delta_x=\Delta_y=33\;{\rm nm}$. b) the same than a) but for a TI ( $\beta=0$ and $\alpha=1\times E_U L_U$) where $E_U=\hbar v_f /Ly$ and $L_U=L_y$. In this case periodic boundary conditions are used for the upper and lower boundaries thus neglecting edge effect.  In this case the interface parameters are:  $J=124.83\;{\rm meV/nm}$, $D=2.5\;{\rm meV/nm^2}$, $\hbar\gamma_0 H_z=0.0004\;{\rm meV}$,  $Jsd=0.1\;{\rm meV}$, $S=1.7$, $v_f=100000\,nm/ns$ and $\mu_L=5\;{\rm meV}$. The resolution of the numerical discretization is: $\Delta_x=\Delta_y=20\;{\rm nm}$.  c)  Torque field resulting from the interaction with a current carried out by a single band of a TI ( $\beta=0$ and $\alpha=1.0\,E_U L_U$) to provide a comparison. The rest of the interface parameters are $J=400\;{\rm meV/nm}$, $D=500\;{\rm meV/nm^2}$, $\hbar\gamma_0 H_z=0.1\;{\rm meV}$,  $Jsd=20.0\;{\rm meV}$, $S=2$ and $\mu_L=75\;{\rm meV}$. Periodic boundary conditions are also used in this case for the upper and lower edges. The resolution of the numerical discretization is: $\Delta_x=\Delta_y=1\;{\rm nm}$.  d), e) and f) Propagating bands in the contacts for a),b) and c) respectively. }
  \label{F9}
\end{figure}

Until now, we have considered conduction electrons in a single conduction band. However, we can also consider the effect on the torque field  of a current flow using many conduction bands. This is shown in fig.\ref{F9} for both the semiconductor and the TI cases considered in this paper. 

In fig.\ref{F9}a we can see the torque field when the current is driven by a semiconductor material through the many bands plotted in fig. \ref{F9}d. The resulting torque field is similar to the one that we obtained from the Zhang et Li classical model for metals\cite{Torque} but tilted because of the resulting effective diagonal spin current. The mean spin current present in this set-up is a combination of a spin-y current flowing from left to right and spin-x current flowing from top to bottom. Furthermore we can see an overall non-zero $-x$ pointing torque perturbed by the skyrmion presence. 

In the central figures \ref{F9}b and \ref{F9}e the same information is shown for the case when the current is driven by a TI. We can also see in fig.\ref{F9}e how the electron dispersion relation in the TI is linear for all of the many bands involved in the transport. Finally, in figs. \ref{F9}c and \ref{F9}f it is possible to find the equivalent plots for the TI when the current is transported by a single band for comparison purposes. Periodic boundaries are also used in this case for the upper and lower edges. Note that the resulting torque field around the skyrmion is similar to both the multiband and the single band transport regimes but it is tilted in the multiband case because in the same way as with the semiconductor material there is a mean diagonal spin current made of a combination of a spin-y left to right current and a spin-x top to bottom current. Interestingly, in a different way than with semiconductor materials a torque field similar to the one provided by the Zhang et Li model is not recovered for the TI in the multiband regime. In the TI case the multiband torque field is closer to the torque field created by current flowing through a single band but with a different orientation of the spatial distribution of the $z$ component of the torque. 

\section{Conclusions}
\label{S6}
The strength of the linear term in a semiconductor or TI material in contact with a ferromagnet has a direct impact on the shape of the torque field
created by the electron current flowing through this material altering the skyrmion movement. This happens because a spin-y accumulation occurs at 
the lower edge of the semiconductor and TI nanowire. In absence of a nearby edge like in wide slabs skyrmion movement by STT is not possible if only a linear term is considered as a model for the material in contact with the ferromagnet. Skyrmion movement is tied to the existence of second order terms when periodic boundary conditions are used on the upper and lower edges of the conducting material.  Furthermore, the torque field it is not fundamentally changed for a  TI when the current is carried by many bands. This means that the classical Zhang et Li model is not applicable for a TI and also that TI's are an inadequate material to induce skyrmion movement due STT unless we exploit edge effects. Further work from a theoretical point of view is needed to determine the efficiency of these materials to create skyrmion movement with SOT instead of STT. The study of SOT driven skyrmion transport is out of scope in our study as this requires a 3d model where a spin current may flow in a perpendicular direction to the ferromagnet-semiconductor/TI interface.

\begin{acknowledgments}
We acknowledge the Horizon 2020 project SKYTOP “Skyrmion-Topological Insulator and Weyl Semimetal Technology”, 
FETPROACT-2018-01, no. 824123. Finally, Javier Osca also acknowledges the postdoctoral fellowship provided by KU Leuven.
This work was supported by IMEC's Industrial Affiliation Program.
\end{acknowledgments} 

\bibliographystyle{apsrev4-1}
\bibliography{skybib}

\end{document}